\documentstyle[11pt]{article}
\begin{document}
\begin{titlepage}
\hfill{UQMATH-93-02}
\hfill{hep-th/9303096}
\vskip.3in
\begin{center}
{\huge Unitarity and Complete Reducibility of Certain Modules
over Quantized Affine Lie Algebras}
\vskip.3in
{\Large Y.-Z.Zhang} and {\Large M.D.Gould}
\vskip.3in
{\large Department of Mathematics, University of Queensland, Brisbane,
Qld 4072, Australia}
\end{center}
\vskip.6in
\begin{center}
{\bf Abstract:}
\end{center}
Let $U_q(\hat{\cal G})$ denote the quantized affine Lie algebra and
$U_q({\cal G}^{(1)})$ the quantized {\em nontwisted} affine Lie algebra.
Let ${\cal O}_{\rm fin}$ be the category defined in section 3.
We show that when the deformation parameter $q$ is not a root of unit
all integrable representations of $U_q(\hat{\cal G})$ in the category
${\cal O}_{\rm fin}$ are completely reducible and that
every integrable irreducible highest weight module over $U_q({\cal G}^{(1)})$
corresponding to $q>0$ is equivalent to a unitary module.

\end{titlepage}

\section{Introduction}
Quantum (super)groups, or more precisely quantum universal enveloping
(super)algebras or for short quantized (super)algebras, are defined
as $q$ deformations of classical
universal enveloping algebras of finite-dimensional simple Lie (super)algebras
\cite{Drinfeld}\cite{Jimbo1}\cite{BGZ}. The definition for
the quantized finite-dimensional simple Lie algebras can be extended to
infinite-dimensional affine Lie algebras,
or even to arbitrary Kac-Moody algebras, with symmetrizable, generalized
Cartan matrices in the sense of Kac\cite{Kac}. In this paper we shall be
concerned with the case of quantized affine Lie algebras.

Quantized affine Lie algebras and their representations are important in,
among others, the so-called Yang-Baxterization method for obtaining spectral
parameter dependent solutions to the quantum Yang-Baxter equation
\cite{Jimbo2}\cite{ZGB} and the
$q$-deformed WZNW CFT's\cite{FR}\cite{Japanese}. It is known\cite{Rosso}
\cite{Lusztig} that most algebras and representations
of interest in physics and mathematics
have corresponding $q$ deformations. In particular,
for quantized simple Lie algebras, all finite-dimensional representations
are known to be completely reducible\cite{Rosso}. However, for quantized
affine Lie algebras, only some isolated results are avaiable. In particular,
the complete reducibility and unitarity of representations have not been
clarified.  In fact the latter remains unproved even for the quantized
finite-dimensional simple Lie algebra case.

In this paper we will address the problems of complete reducibility
and unitarity of certain representations for quantized affine Lie algebras.
Our main results are
the proofs of complete reducibility and unitarity of some important
modules over the quantized affine Lie algebras.

The paper is set up in the following way. After recalling, in section 2,
some basic facts on quantized affine Lie algebras, in section 3 we investigate
representations of the quantized affine Lie algebras. The main result of
this section is theorem 3.1 which says that all integrable representations
of the quantized affine Lie algebras with weight spectrum bounded from
above  are completely
reducible. In section 4 we prove the unitarity of every integrable irreducible
highest weight module over quantized {\em nontwisted} affine Lie algebras;
our main results are stated in theorem 4.2, 4.4 and corollary 4.3. We
conclude, in section 5, with a brief discussion of our main results.

\section{Preliminaries}
\noindent
We start with the definition of the quantum affine Lie
algebra $U_q(\hat{\cal G})$.
Let $A^0=(a_{ij})_{1\leq i,j\leq r}$ be a symmetrizable Cartan matrix.Let
${\cal G}$ stand for the finite-dimensional simple Lie algebra with
symmetric Cartan matrix $A^0_{\rm sym}=(a^{\rm sym}_{ij})=(\alpha_i,
\alpha_j),~~i,j=1,2,...,r$, where $r$ is the rank of ${\cal G}$.
Let $A=(a_{ij})_{0\leq i,j\leq r}$ be a symmetrizable,
generalized Cartan matrix in the sense of Kac. Let $\hat{\cal G}$ denote
the affine Lie algebra associated with the corresponding symmetric Cartan
matrix $A_{\rm sym}=(a^{\rm sym}_{ij})=(\alpha_i,\alpha_j),~~i,j=0,1,...r$.
The quantum algebra $U_q(\hat{\cal G})$ is defined to be a Hopf algebra with
generators: $\{e_i,~f_i,~q^{h_i}~(i=0,1,...,r),~q^d\}$ and relations,
\begin{eqnarray}
&&q^h.q^{h'}=q^{h+h'}~~~~(h,~ h'=h_i~ (i=0,1,...,r),~d)\nonumber\\
&&q^he_iq^{-h}=q^{(h,\alpha_i)} e_i\,,~~q^hf_iq^{-h}=q^{-(h,\alpha_i)}
  f_i\nonumber\\
&&[e_i, f_j]=\delta_{ij}\frac{q^{h_i}-q^{-h_i}}{q-q^{-1}}\nonumber\\
&&\sum^{1-a_{ij}}_{k=0}(-1)^k e_i^{(1-a_{ij}-k)}e_je_i^{(k)}
   =0~~(i\neq j)\nonumber\\
&&\sum^{1-a_{ij}}_{k=0}(-1)^k f_i^{(1-a_{ij}-k)}f_jf_i^{(k)}
   =0~~(i\neq j)\label{relations1}
\end{eqnarray}
where
\begin{equation}
e_i^{(k)}=\frac{e^k_i}{[k]_q!},~~~f^{(k)}_i=\frac{f^k_i}{[k]_q!}
\,,~~~[k]_q=\frac{q^k-q^{-k}}{q-q^{-1}}
\end{equation}

The algebra $U_q(\hat{\cal G})$ is a Hopf algebra with coproduct, counit and
antipode similar to the case of $U_q(\cal G)$:
\begin{eqnarray}
&&\Delta(q^h)=q^h\otimes q^h\,,~~~h=h_i,~d\nonumber\\
&&\Delta(e_i)=q^{-h_i/2}\otimes e_i+e_i\otimes q^{h_i/2}\nonumber\\
&&\Delta(f_i)=
q^{-h_i/2}\otimes f_i+f_i\otimes q^{h_i/2}\nonumber\\
&&S(a)=-q^{h_\rho}aq^{-h_\rho}\,,~~~a=e_i,f_i,h_i,d\label{coproduct1}
\end{eqnarray}
where $\rho$ is the half-sum of the positive roots. We have omitted the
formula for counit since we do not need them.

For quasitriangular Hopf algebras, there exists a distinguished element
\cite{Drinfeld}\cite{KR}
\begin{equation}
u=\sum_iS(b_i)a_i\label{u}
\end{equation}
where $a_i$ and $b_i$ are coordinates of the universal $R$-matrix
$R=\sum_ia_i\otimes b_i$. One can show that $u$ has inverse
\begin{equation}
u^{-1}=\sum_iS^{-2}(b_i)a_i\label{u-1}
\end{equation}
and satisfies
\begin{eqnarray}
&&S^2(a)=uau^{-1}\,,~~~\forall a\in U_q(\hat{\cal G})\nonumber\\
&&\Delta(u)=(u\otimes u)(R^TR)^{-1}
\end{eqnarray}
where $R^T=T(R)$,~~$T$ is the twist map:
$T(a\otimes b)=b\otimes a\,,~\forall a\in U_q(\hat{\cal G})$.
\vskip.1in
\noindent{\bf Proposition 2.1}: $\Omega=uq^{-2h_\rho}$ belongs to the center
of $U_q(\hat{\cal G})$, i.e. it is a Casimir operator.
\vskip.1in
We may equivalently work with the coproduct $\bar{\Delta}$ and antipode
$\bar{S}$ defined by
\begin{eqnarray}
&&\bar{\Delta}(q^h)=q^h\otimes q^h\,,~~h=h_i,~d\nonumber\\
&&\bar{\Delta}(e_i)=q^{h_i/2}\otimes e_i+e_i\otimes q^{-h_i/2}\nonumber\\
&&\bar{\Delta}(f_i)=
q^{h_i/2}\otimes f_i+f_i\otimes q^{-h_i/2}\nonumber\\
&&\bar{S}(a)=-q^{-h_\rho}aq^{h_\rho}\,,~~~a=e_i,f_i,h_i,d\label{coproduct2}
\end{eqnarray}
Corresponding to the coproduct and antipode (\ref{coproduct2})
we have another form of the $R$-matrix,
denoted as $\bar{R}$. If
we write $\bar{R}=\sum_i\bar{a_i}\otimes\bar{b_i}$, then we have
\begin{equation}
\bar{u}=\sum_i\bar{S}(\bar{b_i})\bar{a_i}\,,~~\bar{u}^{-1}=\sum_i
\bar{S}^{-2}(\bar{b_i})\bar{a_i}\label{-u}
\end{equation}
which satisfy
\begin{eqnarray}
&&\bar{S}^2(a)=\bar{u}a\bar{u}^{-1}\,,~~~\forall a\in
  U_q(\hat{\cal G})\nonumber\\
&&\bar{\Delta}(\bar{u})=(\bar{u}\otimes \bar{u})(\bar{R}^T\bar{R})^{-1}
\end{eqnarray}
\vskip.1in
\noindent{\bf Proposition 2.2}: $\bar{\Omega}=\bar{u}^{-1}q^{-2h_\rho}$ is the
Casimir operator of $U_q(\hat{\cal G})$ with coproduct and antipode given
by (\ref{coproduct2}).
\vskip.1in
\noindent{\bf Proposition 2.3}: The Casimir operators $\Omega$ and
$\bar{\Omega}$
have the properties:
$\Omega=S(\Omega)$\,,~$\bar{\Omega}=\bar{S}(\bar{\Omega})$.
\vskip.1in
We define a conjugate operation $\dagger$ and an anti-involution $\theta$ on
$U_q(\hat{\cal G})$ by
\begin{eqnarray}
&&d^\dagger=d\,,~~h_i^\dagger=h_i\,,~~e_i^\dagger=f_i\,,~~f_i^\dagger=e_i
 \,,~~~~i=0,1,...r \nonumber\\
&&\theta(q^h)=q^{-h}\,,~~\theta(e_i)=f_i\,,~~\theta(f_i)=
e_i\,,~~\theta(q)=q^{-1}
\end{eqnarray}
which extend uniquely to an algebra anti-automorphism and anti-involution on
all of $U_q(\hat{\cal G})$, respectively, so that $(ab)^\dagger=b^\dagger
a^\dagger\,,~~\theta(ab)=\theta(b)\theta(a)\,,~~\forall a,b\in U_q(\hat{\cal
G})$.

Throughout the paper, we assume that $q$ {\em is not a root of unit} and
use the notations:
\begin{eqnarray}
&&(n)_q=\frac{1-q^n}{1-q}\,,~~[n]_q=\frac{q^n-q^{-n}}{q-q^{-1}}\,,~~
  q_\alpha=q^{(\alpha,\alpha)}\nonumber\\
&&{\rm exp}_q(x)=\sum_{n\geq 0}\frac{x^n}{(n)_q!}\,,~~(n)_q!=
  (n)_q(n-1)_q\,...\,(1)_q\nonumber\\
&&({\rm ad}_qx_\alpha)x_\beta=[x_\alpha\,,\,x_\beta]_q=x_\alpha x_\beta -
 q^{(\alpha\,,\,\beta)}x_\beta x_\alpha
\end{eqnarray}

\section{Complete Reducibility}
Representations of $U_q(\hat{\cal G})$ are known to be isomorphic (as a linear
space) to corresponding reps of $U(\hat{\cal G})$ \cite{Lusztig}. Let
\begin{equation}
U_q(\hat{\cal G})=
U_q({\cal N}_-)\otimes U_q({\cal H})\otimes U_q({\cal N}_+)\label{decom}
\end{equation}
be the triangular decomposition of $U_q(\hat{\cal G})$, generated
by the $f_i$'s,
$q^h~~(h\in{\cal H})$
and $e_i$'s, respectively.
Let $V$ denote an $U_q(\hat{\cal G})$-module. We say that $V$ is
${\cal H}$-diagonalizable if
\begin{equation}
V=\bigoplus_{\lambda\in {\cal H}^*}V_\lambda\nonumber
\end{equation}
where, $V_\lambda=\{v\in V|q^hv=q^{\lambda
(h)}v\,,h\in {\cal H}\}$ be the weight spaces corresponding to
the weight $\lambda$. Following Kac\cite{Kac}, we introduce the notations
\begin{equation}
\Pi (V)=\{\lambda \in {\cal H}^*|V_\lambda\neq 0\}\,,~~~~
D(\lambda)=\{\mu\in
{\cal H}^*|\mu\leq \lambda\,,~\lambda\in{\cal H}^*\}\nonumber
\end{equation}
and consider the category ${\cal O}_{\rm fin}$ defined by
\vskip.1in
\noindent {\bf Definition 3.1:} ${\cal O}_{\rm fin}$ is the category of $U_q(
\hat{\cal G})$ modules $V$ which are ${\cal H}$-diagonalizable with finite
dimensional weight spaces and in which there exist a finite number of
elements $\lambda_1\,,\,...\,,\lambda_s\,\in\, {\cal H}^*$ such that
\begin{equation}
\Pi (V) \subset \cup^s_{i=1}\, D(\lambda_i)\label{category}
\end{equation}

(\ref{category}) implies that weights in ${\cal O}_{\rm fin}$ are
bounded from above.
\vskip.1in
\noindent\underline{Example:} Highest weight modules.\\
We say that an $U_q(\hat{\cal G})$-module $V$
is a highest weight module with the highest weight $\Lambda$ if there
exists a non-zero vector $v\in V$ such that
\begin{eqnarray}
&&e_i(v)=0~~~(i=0,1,...,r)\,,~~q^hv=q^{\Lambda(h)}v~~(h\in {\cal H})
\nonumber\\
&&V=U_q(\hat{\cal G})v\label{highest}
\end{eqnarray}
The vector $v$ is called as highest weight vector. We deduce from (\ref{decom})
and (\ref{highest}),
\begin{equation}
V=\bigoplus_{\lambda\leq \Lambda}V_\lambda\,,~~~V_\Lambda={\bf C}v\,,~~~
{\rm dim}\;V_\lambda < \infty
\end{equation}
which implies that a highest weight module lies in ${\cal O}_{\rm fin}$.
\vskip.1in
\noindent{\bf Proposition 3.1~(Lusztig \cite{Lusztig}):}
For any $\Lambda\in {\cal H}^*$, there exists a unique, up to isomorphism,
irreducible
highest weight $U_q(\hat{\cal G})$-module $L(\Lambda)$ with highest
weight $\Lambda$.
\vskip.1in
Following Kac\cite{Kac}, we have the following
\vskip.1in
\noindent {\bf Definition 3.2:} Let $V$ be an $U_q(\hat{\cal G})$-module.
A vector
$v\in V_\lambda$ is called primitive of weight $\lambda$ if there exists a
submodule $U$ in $V$ such that
\begin{equation}
v\not\in U\,;\,~~~~~e_iv\in U
\end{equation}
The weight $\lambda$ is called a primitive weight.
\vskip.1in
\noindent \underline{Example:} If $V$ is an irreducible
$U_q(\hat{\cal G})$-module, so that the only proper submodule is
$U=\{0\}$, then a weight vector is
primitive implies that $v\neq 0$ and $e_iv=0$.

We now state two propositions (3.2, 3.3, below) and one lemma
(3.1, below) analogous to Kac's classical results
(compare Kac\cite{Kac}, proposition 9.3, lemma 9.5 and proposition 9.9,
respectively).
\vskip.1in
\noindent {\bf Proposition 3.2:} Let $V$ be a non-zero
$U_q(\hat{\cal G})$-module from
${\cal O}_{\rm fin}$.\\
a)  $V$ contains a non-zero weight vector $v$ such that $e_i(v)=0$; in
particular $V$ contains a primitive vector.\\
b)  The following conditions are equivalent:\\
(i) $V$ is irreducible.\\
(ii) $V$ is a highest weight $U_q(\hat{\cal G})$-module and any primitive
vector of $V$ is a highest weight vector.\\
(iii) $V\simeq L(\Lambda)$ for some $\Lambda\in {\cal H}^*$.
\vskip.1in
The condition (iii) means that the $L(\Lambda)$ exhaust all irreducible modules
from ${\cal O}_{\rm fin}$.
\vskip.1in
\noindent{\bf Lemma 3.1:} Let $V$ be an $U_q(\hat{\cal G})$-module from ${\cal
O
}_{\rm fin}$. If for any two primitive weights $\lambda$ and $\mu$ of $V$,
the inequality $\lambda\geq \mu$ implies $\lambda=\mu$, then the module $V$ is
completely reducible.
\vskip.1in
\noindent {\bf Proof:} The proposition (3.2) and lemma (3.1) are proved exactly
as in Kac (\cite{Kac}, proposition 9.3 and lemma 9.5).~~~$\Box$
\vskip.1in
Let $Q=\sum^r_{i=0}{\bf Z}\alpha_i$ denote the root lattice and set
$Q_+=\sum^r_{i=0}{\bf Z}_+\alpha_i$.
\vskip.1in
\noindent {\bf Proposition 3.3:} a) Let $V(\Lambda)$ be an
$U_q(\hat{\cal G})$-module with highest weight $\Lambda$. If $2(\Lambda+\rho,
\beta)\neq (\beta,\beta)$ for every $\beta\in Q_+$, $\beta\neq 0$, then
$V(\Lambda)$ is irreducible.\\
b) Let $V$ be an $U_q(\hat{\cal G})$-module from ${\cal O}_{\rm fin}$. If
for any two primitive weights $\lambda$ and $\mu$ of $V$, such that
$\lambda-\mu=\beta>0$, one has $2(\lambda+\rho,\beta)\neq (\beta,\beta)$,
then $V$ is completely reducible.
\vskip.1in
\noindent {\bf Proof:} We mimic Kac's proof in the classical case. To this
end, we first prove the following
\vskip.1in
\noindent{\bf Lemma 3.2:} Let $V$ be an $U_q(\hat{\cal G})$-module.\\
a) If there exists $v\in V$ such that $e_iv=0$ for all $i=0,1,...,r$
and $q^hv=q^{\Lambda(h)}v$ for some $\Lambda\in {\cal H}^*$ and all
$h\in {\cal H}$, then
\begin{equation}
\Omega v=q^{-(\Lambda,\Lambda+2\rho)}v
\end{equation}
\noindent b) If, furthermore, $V=U_q(\hat{\cal G})v$, then
\begin{equation}
\Omega|_V=q^{-(\Lambda,\Lambda+2\rho)}I_V
\end{equation}
\vskip.1in
\noindent {\bf Proof:} One can show that the universal $R$-matrix $R$ of
$U_q(\hat{\cal G})$ can be written in the form\cite{FR}
\begin{equation}
R=\left (I\otimes I+\sum_t\,a'_t\otimes b'_t\right )\cdot
q^{\sum^r_{i=1}\,H_i\otimes H^i+
c\otimes d+d\otimes c}
\end{equation}
where $\{a'_t\}$ and $\{b'_t\}$ are the basis of the
subalgebras of $U_q(\hat{\cal
G})$ generated by $\{e_iq^{-h_i/2}\}$ and $\{q^{h_i/2}f_i\}$, $i=0,1,...,r$,
respectively; $c=h_0+h_{\psi}$
\,,~~ $\psi$ is the highest root of
${\cal G}$;~$\{H_i\}$ and $\{H^i\}$~~($i=1,2,...,r$) satisfy
\begin{equation}
\sum^r_{i=1}\Lambda(H_i)\Lambda'(H^i)=(\Lambda_0,\Lambda'_0)\,,~~~~\forall
\Lambda=(\Lambda_0,\kappa,\sigma)\,,\,\Lambda'=(\Lambda'_0,\kappa',\sigma')\in
{\cal H}^*\label{2lambda}
\end{equation}
So the Casimir $\Omega$ in proposition 2.1 takes the form
\begin{equation}
\Omega=q^{-\sum^r_{i=1}\,H^iH_i-dc-cd-2h_\rho}+\sum_t\,S(b'_t)\;
q^{-\sum^r_{i=1}
H^iH_i-dc-cd}\cdot a'_t\;q^{-2h_\rho}
\end{equation}
Acting on $v$, only the first term survives,
\begin{equation}
\Omega v=q^{-\sum^r_{i=1}\,H^iH_i-dc-cd-2h_\rho}v=q^{-(\Lambda,\Lambda+2\rho)}v
\label{eigenvalue}
\end{equation}
where use has been made of (\ref{2lambda}) and
$(\Lambda,\Lambda')=(\Lambda_0,
\Lambda'_0)+\kappa\sigma'+\sigma\kappa'$. This proves a). Part b)
follows from (\ref{eigenvalue}) and  proposition 2.1.~~~~$\Box$
\vskip.1in
\noindent {\bf Corollary 3.1:} a) If $V$ is a highest weight $U_q(\hat{\cal
G})$
module with highest weight $\Lambda$, then
\begin{equation}
\Omega=q^{-(|\Lambda+\rho|^2-|\rho|^2)}I_V
\end{equation}
\noindent b) If $V$ is an $U_q(\hat{\cal G})$-module from ${\cal O}_{\rm fin}$
and $v$ is a primitive vector with weight $\lambda$, then there exists a
submodule $U\subset V$ such that $v\not\in U$ and
\begin{equation}
\Omega v=q^{-(|\Lambda+\rho|^2-|\rho|^2)}v~~({\rm mod}~U)
\end{equation}
\vskip.1in
Now we are in the position to prove proposition 3.3. Assume that
$V(\Lambda)$ is reducible. Then proposition 3.2b) implies that there exists a
primitive weight $\lambda=\Lambda-\beta$, where $\beta>0$ and thus from
corollary 3.1a) we have
\begin{equation}
q^{-(\Lambda,\Lambda+2\rho)}=q^{-(\Lambda-\beta,\Lambda-\beta+2\rho)}
\end{equation}
which gives $2(\Lambda+\rho,\beta)=(\beta,\beta)$ since $q$ is not a root
of unity. This leads to a contradiction and thus we prove a).

We now prove b). We may assume that the $U_q(\hat{\cal G})$-module is
indecomposable. Then, locally, the Casimir operator $\Omega$ has the
same spectrum on $V$. We thus obtain from corollary 3.1b)
\begin{equation}
q^{-(|\lambda+\rho|^2-|\rho|^2)}=q^{-(|\mu+\rho|^2-|\rho|^2)}
\end{equation}
for any two primitive weights $\lambda$ and $\mu$. Since $q$ is not a root
of unity, the above equation gives $|\lambda+\rho|^2=|\mu+\rho|^2$ for any
two primitive weights $\lambda$ and $\mu$. Therefore, we must have
$\lambda=\mu$. Indeed, if this is not the case, then we deduce
$2(\lambda+\rho,\beta)=(\beta,\beta)$, which contradicts the condition of
the proposition. Now point b) follows from lemma 3.1.~~~~$\Box$
\vskip.1in
\noindent {\bf Definition 3.3:} An $U_q(\hat{\cal G})$-module $V$ is called
integrable if $V$ is ${\cal H}$-diagonalizable and if $e_i$ and $f_i$
($i=0,1,...,r$) are locally nilpotent endomorphisms of $V$.
\vskip.1in
Let $\Pi(\Lambda)$ denote the set of weights of the $U_q(\hat{\cal G})$-module
$L(\Lambda)$ and $D_+=\{\lambda\in {\cal H}^*|(\lambda,\alpha_i)\geq 0,~
0\leq i\leq r\}$ the set of dominant integral weights.
\vskip.1in
\noindent {\bf Proposition 3.5:} a) Let $V$ be an $U_q(\hat{\cal G})$-module
from ${\cal O}_{\rm fin}$ and $\lambda$ be a primitive weight. If $V$ is
integrable, then $\lambda\in D_+$.\\
b) The highest weight
$U_q(\hat{\cal G})$-module
$L(\Lambda)$ with highest weight $\Lambda$ is integrable iff $\Lambda\in D_+$.
\vskip.1in
\noindent {\bf Proof:} Part a) is proved following the same arguments
as in Lusztig (\cite{Lusztig}, proposition 3.2) and part b) follows from
\cite{Rosso}\cite{Lusztig}.~~~$\Box$

We now state our main result ({\em complete reducibility} theorem)
in this section.
\vskip.1in
\noindent {\bf Theorem 3.1:} Every integrable $U_q(\hat{\cal G})$-module
$V$ from ${\cal O}_{\rm fin}$ is completely reducible, that is, is
isomorphic to a direct sum of modules $L(\Lambda)$,~
$\Lambda\in D_+$.
\vskip.1in
\noindent {\bf Proof:} We check that if $\lambda$ and $\mu$ are primitive
weights such that $\lambda-\mu=\beta$, where $\beta\in Q_+/\{0\}$, then
\begin{equation}
2(\lambda+\rho,\beta)\neq (\beta,\beta)
\end{equation}
This can easily be done as follows. By means of proposition 3.5a) and
the fact that $(\rho,\beta)>0$ for all $\beta\in Q_+/\{0\}$, we have
\begin{equation}
2(\lambda+\rho,\beta)-(\beta,\beta)=(\lambda+(\lambda-\mu)+2\rho,\beta)
=(\lambda+\mu+2\rho,\beta)>0
\end{equation}
The theorem then follows from proposition 3.3b).~~~~$\Box$

\section{Unitarity}
In this section we will focus our attention on  quantized
{\em nontwisted} affine Lie algebras
$U_q({\cal G}^{(1)})$. Analogous conclusions are true for the {\em twisted}
case. We first introduce the following
\vskip.1in
\noindent{\bf Definition 4.0:}
An $U_q({\cal G}^{(1)})$-module $V$ is called {\em unitary} if $V$
can be equipped with an inner product  $<~|~>$ such that,
for all $a\in U_q({\cal G}^{(1)})$
\begin{equation}
<a^\dagger v|w>=<v|a w>\;,~~~\forall v, w\in V
\end{equation}
Equivalently, if $\pi$ is the representation of $U_q({\cal G}^{(1)})$
afforded by $V$, then $V$ is called unitary provided
\begin{equation}
\pi (a^\dagger)=\pi (a)^\dagger\;,~~~\forall a\in U_q({\cal G}^{(1)})
\end{equation}
where $\dagger$ on the r.h.s. denotes Hermitian conjugate.
\vskip.1in
\noindent {\bf Lemma 4.0:} Every integrable highest weight
$U_q({\cal G}^{(1)})$-module $L(\Lambda)$ carries a unique, up to a constant
factor, and well-defined nondegenerate inner product $<~|~>$. With
respect to this inner product, $L(\Lambda)$ decomposes into an orthogonal
direct
sum of weight spaces.
\vskip.1in
\noindent{\bf Proof:} This can be easily proved following the similar arguments
as in Kac (\cite{Kac}, proposition 9.4).~~~$\Box$
\vskip.1in
\noindent{\bf Proposition 4.0 (Kac)}:
Let $\Lambda\in D_+$ and $\lambda\in\Pi(\Lambda)$. Then
$|\Lambda+\rho|^2-|\lambda+\rho|^2\geq 0$ and equality holds iff $\lambda=
\Lambda$.

\vskip.1in
\noindent 4.1.  Let $\hat{\cal G}={sl(2)}^{(1)}$. Fix a normal ordering in
the positive root system $\Delta_+$ of $sl(2)^{(1)}$:
\begin{equation}
\alpha,\,\alpha+\delta,\,...,\,\alpha+n\delta,\,...,\,\delta,\,2\delta,\,
...,\,m\delta,\,...\,,\,...\,,\,\beta+l\delta,\,...\,,\beta\label{order1}
\end{equation}
where $\alpha$ and $\beta$ are simple roots and $l,m,n\geq 0$;
$\delta=\alpha+\beta$ is the
minimal positive imaginary root. Let us introduce standard generators
\begin{eqnarray}
&&E_\alpha=e_\alpha q^{-h_\alpha/2}\,,~~~~E_\beta=e_\beta q^{-h_\beta/2}
  \nonumber\\
&&F_\alpha=q^{h_\alpha/2}f_\alpha\,,~~~~F_\beta=q^{h_\beta/2}f_\beta
\label{regular1}
\end{eqnarray}
then,
\begin{eqnarray}
&&S(E^\dagger_\alpha)=-F_\alpha\,,~~~~S(E^\dagger_\beta)=-F_\beta\nonumber\\
&&S(F^\dagger_\alpha)=-E_\alpha\,,~~~~S(F^\dagger_\beta)=-E_\beta\label{s1}
\end{eqnarray}
Construct Cartan-Weyl generators $E_\gamma\,,~F_\gamma=\theta(E_\gamma)\,,~~
\gamma\in \Delta_+$ of $U_q(sl(2)^{(1)})$ as follows\cite{KT}:
We define
\begin{eqnarray}
&&\tilde{E_\delta}=[(\alpha,\alpha)]_q^{-1}[E_\alpha,\,E_\beta]_q\nonumber\\
&&E_{\alpha+n\delta}=(-1)^n\left ({\rm ad}\tilde{E_\delta}\right )^nE_\alpha
  \nonumber\\
&&E_{\beta+n\delta}=\left ({\rm ad}\tilde{E_\delta}\right )^nE_\beta\,,...
  \nonumber\\
&&\tilde{E}_{n\delta}=(\alpha,\alpha)]_q^{-1}[E_{\alpha+(n-1)\delta},\,
E_\beta]_q  \label{cartan-weyl1}
\end{eqnarray}
where $[\tilde{E}_{n\delta},\,\tilde{E}_{m\delta}]=0$ for any $n,\,m >0$. For
any $n>0$ there exists a unique element $E_{n\delta}$ \cite{KT} which satisfies
$[E_{n\delta}\,,\,E_{m\delta}]=0$ for any $n,\,m>0$ and the relation
\begin{equation}
\tilde{E}_{n\delta}=\sum_{
\begin{array}{c}
k_1p_1+...+k_mp_m=n\\
0<k_1<...<k_m
\end{array}
}\frac{\left ( q^{(\alpha,\alpha)}-q^{-(\alpha,\alpha)}\right )^{\sum_ip_i-1}}
{p_1!\;...\;p_m!}(E_{k_1\delta})^{p_1}...(E_{k_m\delta})^{p_m} \label{ee1}
\end{equation}
Then the vectors $E_\gamma$ and $F_\gamma=
\theta(E_\gamma)$, $\gamma\in \Delta_+$  defined above
are the Cartan-Weyl generators for
$U_q({sl(2)}^{(1)})$. Moreover,
\vskip.1in
\noindent {\bf Theorem 4.1} (Khoroshkin-Tolstoy\cite{KT}): The universal
$R$-matrix for $U_q(sl(2)^{(1)})$ may be written as
\begin{eqnarray}
R&=&\left ( \Pi_{n\geq 0}{\rm exp}_{q_\alpha}((q-q^{-1})(E_{\alpha+n\delta}
  \otimes F_{\alpha+n\delta}))\right )\nonumber\\
  & &\cdot{\rm exp}\left ( \sum_{n>0}n[n]_{q_\alpha}^{-1}
  (q_\alpha-q_\alpha^{-1})(E_{n\delta}\otimes F_{n\delta})\right )\nonumber\\
& &\cdot\left (\Pi_{n\geq 0}{\rm exp}_{q_\alpha}((q-q^{-1})(E_{\beta+n\delta}
  \otimes F_{\beta+n\delta}))\right )\cdot
  q^{\frac{1}{2}h_\alpha\otimes h_\alpha+c\otimes d+d\otimes c}\label{sl2R}
\end{eqnarray}
where $c=h_\alpha+h_\beta$. The order in the product (\ref{sl2R}) concides
with the chosen normal order (\ref{order1}).

We have
\vskip.1in
\noindent {\bf Lemma 4.1:}
\begin{eqnarray}
&&S(E^\dagger_{\alpha+n\delta})=-q^{n(\alpha,\beta)}F_{\alpha+n\delta}\,,~~~
  S(E^\dagger_{\beta+n\delta})=-q^{n(\alpha,\beta)}F_{\beta+n\delta}\nonumber\\
&&S(F^\dagger_{\alpha+n\delta})=-q^{-n(\alpha,\beta)}E_{\alpha+n\delta}\,,~~~
  S(F^\dagger_{\beta+n\delta})=-q^{-n(\alpha,\beta)}
  E_{\beta+n\delta}\nonumber\\
&&S(\tilde{E}^\dagger_{n\delta})=-q^{n(\alpha,\beta)}\tilde{F}_{n\delta}\,,~~~
  S(E^\dagger_{n\delta})=-q^{n(\alpha,\beta)}F_{n\delta}\nonumber\\
&&S(\tilde{F}^\dagger_{n\delta})=-q^{-n(\alpha,\beta)}\tilde{E}_{n\delta}\,,~~~
  S(F^\dagger_{n\delta})=-q^{-n(\alpha,\beta)}E_{n\delta}\label{s2}
\end{eqnarray}
\vskip.1in
\noindent {\bf Proof:} The proof follows, from (\ref{s1}), (\ref{cartan-weyl1})
and (\ref{ee1}) and similar relations for $F_\gamma=\theta(E_\gamma)$, by
induction in $n$.~~~$\Box$
\vskip.1in
\noindent {\bf Corollary 4.1:}
\begin{eqnarray}
&&S(F_{\alpha+n\delta})=-q^{-n(\alpha,\beta)}S^2(E^\dagger_{\alpha+n\delta})
  =-q^{-n(\alpha,\beta)-(\alpha+n\delta,2\rho)}E^\dagger_{\alpha+n\delta}
  \nonumber\\
&&S(F_{\beta+n\delta})=-q^{-n(\alpha,\beta)}S^2(E^\dagger_{\beta+n\delta})
  =-q^{-n(\alpha,\beta)-(\beta+n\delta,2\rho)}E^\dagger_{\beta+n\delta}
  \nonumber\\
&&S(F_{n\delta})=-q^{-n(\alpha,\beta)}S^2(E^\dagger_{n\delta})
  =-q^{-n(\alpha,\beta)-(n\delta,2\rho)}E^\dagger_{n\delta}
\end{eqnarray}

We are now ready to state
\vskip.1in
\noindent {\bf Theorem 4.2:} Every integrable highest weight module
$L(\Lambda)$ over $U_q(sl(2)^{(1)})$  corresponding to $q>0$ is equivalent
to a unitary module.
\vskip.1in
\noindent {\bf Proof:} In the limit $q\rightarrow 1$, the
$U_q({sl(2)}^{(1)})$-module
$L(\Lambda)$ reduces to the corresponding module of $U({sl(2)}^{(1)})$
\cite{Lusztig} and thus is
equivalent to a unitary module according to Kac\cite{Kac}.
We now show that for $0<q<1$ and $q>1$ the
module $L(\Lambda)$ is equivalent to a unitary module. By lemma 4.0,
one only need to show that
if $<~|~>$ is a nondegenerate inner product on $L(\Lambda)$ such that
$<v|v>~~>~0$ for a highest weight vector $v$, then the restriction of $<~|~>$
to $L(\Lambda)_\lambda$ is positive definite for each weight $\lambda$ in
$L(\Lambda)$. We prove this by induction on
${\rm ht}(\Lambda-\lambda)$ (the height of $(\Lambda-\lambda)$).
Let $\lambda\in\Pi(\Lambda)/\{\Lambda\}$.\\
(i). \underline{For $0<q<1$}:\\
we use the Casimir $\Omega=uq^{-2h_\rho}$. We have, from the $R$-matrix
(\ref{sl2R}),
\begin{eqnarray}
u&=&\sum_{\{{\bf l},{\bf n},{\bf k}\}}\;A_{{\bf l},
  {\bf n},{\bf k}}(q)\,S(F_\beta)^{k_0}\,...\,S(F_{\beta+M\delta})^{k_M}
 \,...\;...\, S(F_{L\delta})^{n_L}\,...\,S(F_\delta)^{n_1}\nonumber\\
  & &\cdot \,...\,S(F_{\alpha+N\delta})^{l_N}\,
  ...\,S(F_\alpha)^{l_0}
 q^{-\frac{1}{2}h_\alpha h_\alpha-cd-dc}(E_\alpha)^{l_0}\,...\,
  (E_{\alpha+N\delta})^{l_N}\,...\, \nonumber\\
  & &\cdot (E_\delta)^{n_1}\,...\,(E_{L\delta})^{n_L}\,...\;...\,
  (E_{\beta+M\delta})^{k_M}\,...\,(E_\beta)^{k_0}
\end{eqnarray}
where $\{{\bf l}\}=\{l_0,l_1,...,l_N,...\},~~\{{\bf n}\}=\{n_1,n_2,...,n_L,...
\}$\,,~~$\{{\bf k}\}=\{k_0,k_1,...,k_M,...\}$;~
the constants $A_{{\bf l},{\bf n},{\bf k}}(q)$ are given by
\begin{eqnarray}
A_{{\bf l},{\bf n},{\bf k}}&=&\frac{(q-q^{-1})^{l_0+l_1+...+l_N+...}}
{(l_0)_{q_\alpha}!\,...\,(l_N)_{q_\alpha}!\,...}
\frac{(q-q^{-1})^{k_0+k_1+...+k_M+...}}
{(k_0)_{q_\alpha}!\,...\,(k_M)_{q_\alpha}!\,...}\nonumber\\
& &\cdot \frac{1^{n_1}\,...\,L^{n_L}\,...\;(q-q^{-1})^{n_1+n_2+...+n_L+...}}
{[1]_{q_\alpha}^{n_1}\,...\,[L]_{q_\alpha}^{n_L}\,...\;n_1!\,...\,n_L!\,...}
\end{eqnarray}
and satisfy
\begin{equation}
(-1)^{l_0+...+l_N+...}(-1)^{n_1+...+n_L+...}(-1)^{k_0+...+k_M+...}
A_{{\bf l},{\bf n},{\bf k}}
(q) > 0\,~~~~~{\rm for}~~ 0<q<1\label{0<q<1}
\end{equation}
Then the corollary 4.1 implies that
\begin{eqnarray}
\Omega&=&q^{-\frac{1}{2}h_\alpha h_\alpha-cd-dc-2h_\rho}+
  \sum_{\{{\bf l},{\bf n},{\bf k}\}}'\;
 \hat{A}_{{\bf l},{\bf n},{\bf k}}(q)\,(E^\dagger_\beta)^{k_0}\,...\,
 (E^\dagger_{\beta+M\delta})^{k_M}\,...\;...\,\nonumber\\
  & &\cdot (E^\dagger_{L\delta})^{n_L}\,...\,(E^\dagger_\delta)^{n_1}
  \,...\,(E^\dagger_{\alpha+N\delta})^{l_N}\,
  ...\,(E^\dagger_\alpha)^{l_0}
  q^{-\frac{1}{2}h_\alpha h_\alpha-cd-dc}
 (E_\alpha)^{l_0}\,...\,
  (E_{\alpha+N\delta})^{l_N}\nonumber\\
  & &\cdot \,...\,(E_\delta)^{n_1}\,...\,(E_{L\delta})^{n_L}\,...\;...\,
  (E_{\beta+M\delta})^{k_M}\,...\,(E_\beta)^{k_0}q^{-2h_\rho}
\end{eqnarray}
where $\hat{A}_{{\bf l},{\bf n},{\bf k}}(q)=(-1)^{l_0+...+l_N+...}
(-1)^{n_1+...+n_L+...}(-1)^{k_0+...+k_M+...}
A_{{\bf l},{\bf n},{\bf k}}(q)$ and so by (\ref{0<q<1})
\begin{equation}
\hat{A}_{{\bf l},{\bf n},{\bf k}}(q)>0\,~~~~~{\rm for}~~~0<q<1\label{a>0}
\end{equation}
and $\sum'$ denotes the sum over all $\{{\bf l},{\bf n},{\bf k}\}\neq
\{{\bf 0},{\bf 0},{\bf 0}\}$.

Computing $<v|\Omega|v>$ in two different ways, we obtain
\begin{eqnarray}
&&\left( q^{-(|\Lambda+\rho|^2-|\rho|^2)}-q^{-(|\lambda+\rho|^2-|\rho|^2)}
\right ) <v|v>=\sum_{\{{\bf l},{\bf n},{\bf k}\}}'\;
  \hat{A}_{{\bf l},{\bf n},{\bf k}}(q)\,q^{-2(\lambda,\rho)}
\cdot <  (E_\alpha)^{l_0}\,...\,
  (E_{\alpha+N\delta})^{l_N}\,...\,\nonumber\\
&&~~~~~~\cdot  (E_\delta)^{n_1}\,...\,(E_{L\delta})^{n_L}\,...\;...\,
  (E_{\beta+M\delta})^{k_M}\,...\,(E_\beta)^{k_0}v|
  q^{-\frac{1}{2}h_\alpha h_\alpha-cd-dc}\nonumber\\
&&~~~~~~\cdot (E_\alpha)^{l_0}\,...\,
  (E_{\alpha+N\delta})^{l_N}\,...\,(E_\delta)^{n_1}\,...\,(E_{L\delta})^{n_L}
  \,...\;...\,(E_{\beta+M\delta})^{k_M}\,...\,(E_\beta)^{k_0}v>\label{1}
\end{eqnarray}
By the inductive assumption the r.h.s. of (\ref{1}) is non-negative thanks to
eq.(\ref{a>0}). Using proposition 4.0 we deduce that $<v|v>~~\geq~ 0$
for       $0<q<1$.
Since $<~|~>$ is non-degenerate on $L(\Lambda)_\lambda$ we conclude that
for $0<q<1$ it
is positive definite on $L(\Lambda)$.\\
(ii). \underline{For $q>1$}:\\
In this case we work with the coproduct and
antipode (\ref{coproduct2}). Let us introduce the standard generators
\begin{eqnarray}
&&\bar{E}_\alpha=e_\alpha q^{h_\alpha/2}\,,~~~~\bar{E}_\beta=e_\beta
  q^{h_\beta/2}\nonumber\\
&&\bar{F}_\alpha=q^{-h_\alpha/2}f_\alpha\,,~~~~\bar{F}_\beta=
  q^{-h_\beta/2}f_\beta\label{regular2}
\end{eqnarray}
then
\begin{eqnarray}
&&\bar{S}(\bar{E}^\dagger_\alpha)=-\bar{F}_\alpha\,,~~~~\bar{S}
  (\bar{E}^\dagger_\beta)=-\bar{F}_\beta\nonumber\\
&&\bar{S}(\bar{F}^\dagger_\alpha)=-\bar{E}_\alpha\,,~~~~\bar{S}
  (\bar{F}^\dagger_\beta)=-\bar{E}_\beta
\end{eqnarray}
and we have
\vskip.1in
\noindent {\bf Lemma 4.2:}
\begin{eqnarray}
&&\bar{S}(\bar{F}_{\alpha+n\delta})
 =-q^{n(\alpha,\beta)+(\alpha+n\delta,2\rho)}\bar{E}^\dagger_{\alpha+n\delta}
  \nonumber\\
&&\bar{S}(\bar{F}_{\beta+n\delta})
  =-q^{n(\alpha,\beta)+(\beta+n\delta,2\rho)}\bar{E}^\dagger_{\beta+n\delta}
  \nonumber\\
&&\bar{S}(\bar{F}_{n\delta})
  =-q^{n(\alpha,\beta)+(n\delta,2\rho)}\bar{E}^\dagger_{n\delta}
\end{eqnarray}
\vskip.1in
\noindent {\bf Proof:} Similar to the proof of lemma 4.1.~~~$\Box$
\vskip.1in
Now we define inductively,
\begin{eqnarray}
&&\tilde{{\bar E}_\delta}=[(\alpha,\alpha)]_q^{-1}[\bar{E}_\alpha,\,
  \bar{E}_\beta]_{q^{-1}}\nonumber\\
&&\bar{E}_{\alpha+n\delta}=(-1)^n\left ({\rm ad}\tilde{{\bar E}_\delta}
  \right )^n\bar{E}_\alpha\nonumber\\
&&\bar{E}_{\beta+n\delta}=\left ({\rm ad}\tilde{\bar{E}_\delta}\right )^n
  \bar{E}_\beta\,,...\nonumber\\
&&\tilde{\bar E}_{n\delta}=[(\alpha,\alpha)]_q^{-1}
  [\bar{E}_{\alpha+(n-1)\delta},\,\bar{E}_\beta]_{q^{-1}}\nonumber\\
&&\tilde{\bar E}_{n\delta}=\sum_{
\begin{array}{c}
k_1p_1+...+k_mp_m=n\\
0<k_1<...<k_m
\end{array}
}\frac{\left ( q^{-(\alpha,\alpha)}-q^{(\alpha,\alpha)}\right )^{\sum_ip_i-1}
}{p_1!\;...\;p_m!}(\bar{E}_{k_1\delta})^{p_1}...
(\bar{E}_{k_m\delta})^{p_m}\label{ee1'}
\end{eqnarray}
and similarly for $\bar{F}_\gamma=\theta(\bar{E}_\gamma)$. Then we immediately
obtain, from the $R$-matrix (\ref{sl2R}), our matrix
$(\bar{R}^{\rm T})^{-1}$ :
\begin{eqnarray}
(\bar{R}^{\rm T})^{-1}&=&
\sum_{\{{\bf l},{\bf n},{\bf k}\}}\;A_{{\bf l},
  {\bf n},{\bf k}}(q^{-1})\,
(\bar{E}_\alpha)^{l_0}\,...\,
(\bar{E}_{\alpha+N\delta})^{l_N}\,...\,(\bar{E}_\delta)^{n_1}\,...\,
(\bar{E}_{L\delta})^{n_L}\,...\,\nonumber\\
& &\cdot \,...\,(\bar{E}_{\beta+M\delta})^{k_M}\,...\,(\bar{E}_\beta)^{k_0}
 \otimes (\bar{F}_\alpha)^{l_0}\,...\,
  (\bar{F}_{\alpha+N\delta})^{l_N}\,...\,
  (\bar{F}_\delta)^{n_1}\,...\,(\bar{F}_{L\delta})^{n_L}\nonumber\\
  & &\cdot \,...\;...\,(\bar{F}_{\beta+M\delta})^{k_M}\,...\,
  (\bar{F}_\beta)^{k_0}
 \cdot q^{-\frac{1}{2}h_\alpha\otimes h_\alpha-c\otimes d-d\otimes c}\label{2}
\end{eqnarray}
where the constants $A_{{\bf l},{\bf n},{\bf k}}(q^{-1})$ satisfy
\begin{equation}
(-1)^{l_0+...+l_N+...}(-1)^{n_1+...+n_L+...}(-1)^{k_0+...+k_M+...}
A_{{\bf l},{\bf n},{\bf k}}(q^{-1})>0\,,~~~~~~{\rm for}~~~ q>1\label{q>1'}
\end{equation}
We deduce, from (\ref{2}),
\begin{eqnarray}
\bar{R}&=&(I\otimes \bar{S})\bar{R}^{-1}=
\sum_{\{{\bf l},{\bf n},{\bf k}\}}\;A_{{\bf l},
  {\bf n},{\bf k}}(q^{-1})\,
(\bar{F}_\alpha)^{l_0}\,...\,
(\bar{F}_{\alpha+N\delta})^{l_N}\,...\,
(\bar{F}_\delta)^{n_1}\,...\,(\bar{F}_{L\delta})^{n_L}\,...\,\nonumber\\
& &\cdot \,...\, (\bar{F}_{\beta+M\delta})^{k_M}\,...\,(\bar{F}_\beta)^{k_0}
 \otimes q^{\frac{1}{2}h_\alpha\otimes h_\alpha+c\otimes d+d\otimes c}
\cdot \bar{S}(\bar{E}_\beta)^{k_0}\,...\,
 \bar{S}(\bar{E}_{\beta+M\delta})^{k_M}\,...\,\nonumber\\
& &\cdot \,...\, \bar{S}(\bar{E}_{L\delta})^{n_L}\,...\,
\bar{S}(\bar{E}_{\delta})^{n_1}\,...\,
 \bar{S}(\bar{E}_{\alpha+N\delta})^{l_N}\,...\,\bar{S}(\bar{E}_\alpha)^{l_0}
\end{eqnarray}
Thus we obtain the following Casimir operator
\begin{eqnarray}
\bar{\Omega}&=&\bar{S}(\bar{\Omega})=\bar{S}(\bar{u}^{-1}q^{-2h_\rho})=
  q^{2h_\rho}\bar{S}(\bar{u}^{-1})\nonumber\\
& &=q^{2h_\rho}\sum_{\{{\bf l},{\bf n},{\bf k}\}}
  A_{{\bf l},{\bf n},{\bf k}}(q^{-1})\,
\bar{S}(\bar{F}_\beta)^{k_0}\,...\,
\bar{S}(\bar{F}_{\beta+M\delta})^{k_M}\,...\;...\,\nonumber\\
& &\cdot \bar{S}(\bar{F}_{L\delta})^{n_L}\,...\,\bar{S}(\bar{F}_{\delta})^{n_1}
 \,...\,\bar{S}(\bar{F}_{\alpha+N\delta})^{l_N}\,...\,
 \bar{S}(\bar{F}_\alpha)^{l_0}
\cdot q^{\frac{1}{2}h_\alpha h_\alpha+c d+d c}\nonumber\\
& &\cdot (\bar{E}_\alpha)^{l_0}\,...\,
 (\bar{E}_{\alpha+N\delta})^{l_N}\,...\,
 (\bar{E}_{\delta})^{n_1}\,...\,(\bar{E}_{L\delta})^{n_L}\,...\;...\,
 (\bar{E}_{\beta+M\delta})^{k_M}\,...\,(\bar{E}_\beta)^{k_0}
\end{eqnarray}
which, using the lemma 4.2, takes the form
\begin{eqnarray}
\bar{\Omega}&=&q^{\frac{1}{2}h_\alpha h_\alpha+cd+dc+2h_\rho}+
  \sum_{\{{\bf l},{\bf n},{\bf k}\}}'\;
 \hat{A}_{{\bf l},{\bf n},{\bf
k}}(q^{-1})\,(\bar{E}^\dagger_\beta)^{k_0}\,...\,
 (\bar{E}^\dagger_{\beta+M\delta})^{k_M}\,...\;...\,\nonumber\\
& &\cdot (\bar{E}^\dagger_{L\delta})^{n_L}\,...\,(\bar{E}^\dagger_\delta)^{n_1}
  \,...\,(\bar{E}^\dagger_{\alpha+N\delta})^{l_N}\,
  ...\,(\bar{E}^\dagger_\alpha)^{l_0}
  q^{\frac{1}{2}h_\alpha h_\alpha+cd+dc}
 (\bar{E}_\alpha)^{l_0}\,...\,
  (\bar{E}_{\alpha+N\delta})^{l_N}\nonumber\\
  & &\cdot \,...\,(\bar{E}_\delta)^{n_1}\,...\,(\bar{E}_{L\delta})^{n_L}
  \,...\;...\,(\bar{E}_{\beta+M\delta})^{k_M}\,...\,(\bar{E}_\beta)^{k_0}
\end{eqnarray}
where $\hat{A}_{{\bf l},{\bf n},{\bf k}}(q^{-1})=(-1)^{l_0+...+l_N+...}
(-1)^{n_1+...+n_L+...}(-1)^{k_0+...+k_M+...}
A_{{\bf l},{\bf n},{\bf k}}(q^{-1})$ and so by (\ref{q>1'})
\begin{equation}
\hat{A}_{{\bf l},{\bf n},{\bf k}}(q^{-1})>0\,~~~~~{\rm for}~~~q>1\label{q>1}
\end{equation}
Computing $<v|\bar{\Omega}|v>$ in two different ways as above, we obtain
\begin{eqnarray}
&&\left( q^{|\Lambda+\rho|^2-|\rho|^2}-q^{|\lambda+\rho|^2-|\rho|^2}
\right ) <v|v>=\sum_{\{{\bf l},{\bf n},{\bf k}\}}'\;
  \hat{A}_{{\bf l},{\bf n},{\bf k}}(q^{-1})\,q^{2(\lambda,\rho)}
  \cdot< (\bar{E}_\alpha)^{l_0}\,...\,
  (\bar{E}_{\alpha+N\delta})^{l_N}\,...\,\nonumber\\
&&~~~~~~\cdot  (\bar{E}_\delta)^{n_1}\,...\,(\bar{E}_{L\delta})^{n_L}
  \,...\;...\,(\bar{E}_{\beta+M\delta})^{k_M}\,...\,(\bar{E}_\beta)^{k_0}v|
  q^{\frac{1}{2}h_\alpha h_\alpha+cd+dc}\nonumber\\
&&~~~~~~\cdot (\bar{E}_\alpha)^{l_0}\,...\,
  (\bar{E}_{\alpha+N\delta})^{l_N}\,...\,(\bar{E}_\delta)^{n_1}\,...\,
  (\bar{E}_{L\delta})^{n_L}\,...\;...\,
  (\bar{E}_{\beta+M\delta})^{k_M}\,...\,(\bar{E}_\beta)^{k_0}v>\label{3}
\end{eqnarray}
By the inductive hypothesis we have that the r.h.s. of (\ref{3}) is
non-negative for $q>1$ thanks to the formula (\ref{q>1}). Therefore, we
deduce from  proposition 4.0 that $<v|v>~~\geq ~0$ for $q>1$. Then the
non-degeneracy of $<|>$ on $L(\Lambda)$ implies that $<v|v>~~>0$ for
$q>1$.~~~~$\Box$

\vskip.2in
\noindent 4.2. General case: $\hat{\cal G}={\cal G}^{(1)}$.
Fix some order in the positive
root system $\Delta_+$ of ${\cal G}^{(1)}$, which satisfies
an additional condition,
\begin{equation}
\alpha+n\delta~\leq~k\delta~\leq~(\delta-\beta)+l\delta\label{order2}
\end{equation}
where $\alpha\,,~\beta\,\in~\Delta^0_+$\,,~~$\Delta_+^0$ is the positive
root system of ${\cal G}$\,;~$k\,,\,l\,,\,n\,\geq\,0$ and
$\delta$ is the minimal positive imaginary root.

Let us as before introduce standard generators,
\begin{equation}
E_i=e_iq^{-h_i/2}\,,~~~~F_i=q^{h_i/2}f_i\,,~~~~i=0,1,...,r
\end{equation}
then we have
\begin{equation}
S(E_i^\dagger)=-F_i\,,~~~~S(F_i^\dagger)=-E_i\label{ad1}
\end{equation}
Cartan-Weyl generators $E_\gamma$ and $F_\gamma=\theta(E_\gamma)\,,~~\gamma
\in \Delta_+$ may be constructed inductively as follows\cite{KT}.
We start from the simple
roots. If $\gamma=\alpha+\beta\,,~~\alpha<\gamma<\beta$, is a root and there
are
no other positive roots $\alpha'$ and $\beta'$ between $\alpha$ and $\beta$
such that $\gamma=\alpha'+\beta'$, then we set
\begin{equation}
E_\gamma=[E_\alpha\,,\,E_\beta]_q=E_\alpha E_\beta - q^{(\alpha,\beta)}E_\beta
 E_\alpha
\end{equation}
When we get the root $\delta$, we use the following formula for roots
$\gamma+n\delta$ and roots $(\delta-\gamma)+n\delta$, for
$\gamma\in\Delta_+^0$,
\begin{eqnarray}
&&\tilde{E}_\delta^{(i)}=[(\alpha_i,\alpha_i)]_q^{-1}[E_{\alpha_i},\,
E_{\delta-\alpha_i}]_q\,,\nonumber\\
&&E_{\alpha_i+n\delta}=(-1)^n\left ({\rm ad}\tilde{E}_\delta^{(i)}\right )^n
E_{\alpha_i}\,,\nonumber\\
&&E_{\delta-\alpha_i+n\delta}=\left ({\rm ad}\tilde{E}_\delta^{(i)}\right )^n
E_{\delta-\alpha_i}\,,~~~~  ...~~~,\nonumber\\
&&\tilde{E}_{n\delta}^{(i)}= [(\alpha_i,\alpha_i)]_q^{-1}[E_{\alpha_i
+(n-1)\delta},\,E_{\delta-\alpha_i}]_q  \label{cartan-weyl3}
\end{eqnarray}
Then we repeat the above inductive proceduce to obtain other real root vectors
$E_{\gamma+n\delta}\,,~~E_{\delta-\gamma+n\delta}$\,,
$~\gamma\in\Delta_+^0$. Finally,
the imaginary root vectors $E^{(i)}_{n\delta}$ are defined through
$\tilde{E}^{(i)}_{n\delta}$ by the relation (\ref{ee1}) with $\alpha$ there
changing to $\alpha_i$. Then,
the above operators $E^{(i)}_{n\delta}\,,~~
F^{(i)}_{n\delta}=\theta(E^{(i)}_{n\delta})~~(i=1,2,...,r)\,,~E_\gamma\,,~~
F_\gamma=\theta(E_\gamma)$ are the Cartan-Weyl generators of
$U_q({\cal G}^{(1)})$. Moreover
\vskip.1in
\noindent {\bf Theorem 4.3} (Khoroshkin-Tolstoy\cite{KT}): The universal
$R$-matrix $U_q({\cal G}^{(1)})$
may be written in the following form,
\begin{eqnarray}
R&=&\left (\Pi_{\gamma\in \Delta_+^{\rm re}\,,\,\gamma<\delta}~~{\rm exp}_{q_
\gamma}\left (\frac{q-q^{-1}}{C_\gamma(q)}E_\gamma\otimes F_\gamma\right )
\right )\nonumber\\
& &\cdot {\rm exp}\left (\sum_{n>0}\sum^r_{i,j=1}
C^n_{ij}(q)(q-q^{-1})(E^{(i)}_{n\delta}\otimes F^{(j)}_{n\delta})
\right )\nonumber\\
& &\cdot \left (\Pi_{\gamma\in \Delta_+^{\rm re}\,,\,\gamma>\delta}~~
{\rm exp}_{q_
\gamma}\left (\frac{q-q^{-1}}{C_\gamma(q)}E_\gamma\otimes F_\gamma\right )
\right )\cdot q^{\sum^r_{i,j=1}\,(a^{-1}_{\rm sym})^{ij}h_i\otimes h_j
+c\otimes d+d\otimes c}\label{generalR}
\end{eqnarray}
where $c=h_0+h_{\psi}$,~$\psi$ is the highest root of ${\cal G}$;
$(C^n_{ij}(q))=(C^n_{ji}(q))\,,~~i,j=1,2,...,r$, is the inverse of the matrix
$(B^n_{ij}(q))\,,~~i,j=1,2,...,r$ with
\begin{equation}
B^n_{ij}(q)=(-1)^{n(1-\delta_{ij})}n^{-1}\frac{q^n_{ij}-q^{-n}_{ij}}
{q_{j}-q^{-1}_{j}}\frac{q-q^{-1}}{q_{i}-q^{-1}_{i}}
\,,~~~~q_{ij}=q^{(\alpha_i,\alpha_j)}\,,~~~q_i\equiv q_{\alpha_i}
\end{equation}
and $C_\gamma(q)$ is a normalizing constant defined by
\begin{equation}
[E_\gamma\,,\,F_\gamma]=\frac{C_\gamma(q)}{q-q^{-1}}\left ( q^{h_\gamma}
-q^{-h_\gamma}\right )\,,~~~~\gamma\in\Delta^{\rm re}_+
\end{equation}
The order in the product of the $R$-matrix concides with the chosen normal
ordering (\ref{order2}) in $\Delta_+$. We now state an important
\vskip.1in
\noindent {\bf Remark:}
$C_\gamma(q)$ have the following general property
\begin{equation}
C_\gamma(q)=C_\gamma(q^{-1})~>~0\,~~~~{\rm for}~~q~>~0\,,~~ q~\neq~1
\label{remark}
\end{equation}
as shown in our previous paper\cite{ZG}.

We have the following
\vskip.1in
\noindent {\bf Lemma 4.3:} For any $\alpha\in \Delta^0_+$,
\begin{equation}
S(E^\dagger_\alpha)=-q^{(\alpha,\alpha-2\rho)/2}F_\alpha\,,~~~~~
S(F^\dagger_\alpha)=-q^{-(\alpha,\alpha-2\rho)/2}E_\alpha
\end{equation}
\vskip.1in
\noindent {\bf Proof:} We prove them by induction. The results obviously
are valid for $\alpha=\alpha_i\,,~~i=1,2,...,r$, a simple root since we have
$(\alpha_i,\alpha_i-2\rho)=0$. Now we show that the results are also
true for $E_{\alpha+\beta}=[E_\alpha\,,\,E_\beta]_q$ and $F_{\alpha+\beta}=
[F_\beta\,,\,F_\alpha]_{q^{-1}}$. We have
\begin{equation}
S(E^\dagger_{\alpha+\beta})=S(E^\dagger_\alpha)S(E^\dagger_\beta)
-q^{(\alpha,\beta)}S(E^\dagger_\beta)S(E^\dagger_\alpha)
\end{equation}
which by the inductive assumption gives
\begin{eqnarray}
S(E^\dagger_{\alpha+\beta})&=&q^{(\alpha,\alpha-2\rho)/2+(\beta,\beta-2\rho)/2}
\left( F_\alpha F_\beta-q^{(\alpha,\beta)}F_\beta F_\alpha\right )\nonumber\\
&=&q^{(\alpha,\beta)+(\alpha,\alpha-2\rho)/2+(\beta,\beta-2\rho)/2}
\left( F_\beta F_\alpha-q^{-(\alpha,\beta)}F_\alpha F_\beta\right )\nonumber\\
&=&-q^{(\alpha+\beta,\alpha+\beta-2\rho)/2}F_{\alpha+\beta}
\end{eqnarray}
Similarly, we have
\begin{equation}
S(F^\dagger_{\alpha+\beta})=-q^{-(\alpha+\beta,\alpha+\beta-2\rho)/2}
E_{\alpha+\beta}
\end{equation}
This completes our proof.~~~$\Box$
\vskip.1in
\noindent {\bf Corollary 4.2:} For any $\alpha\in \Delta^0_+$,
\begin{equation}
S(E^\dagger_{\delta-\alpha})=-q^{(\delta-\alpha,\delta-\alpha-2\rho)/2}
F_{\delta-\alpha}\,,~~~~
S(F^\dagger_{\delta-\alpha})=-q^{-(\delta-\alpha,\delta-\alpha-2\rho)/2}
E_{\delta-\alpha}
\end{equation}
\vskip.1in
\noindent {\bf Proof:} The results are true for $\alpha=\psi$, the highest
root. Then the results follow, from lemma 4.3 and $E_{\delta-\alpha}=
[E_\beta\,,\,E_{\delta-(\alpha+\beta)}]_q$ and $F_{\delta-\alpha}=[
F_{\delta-(\alpha+\beta)}\,,\,F_\beta]_{q^{-1}}$, by induction as above.~~~
$\Box$
\vskip.1in
\noindent {\bf Lemma 4.4:} For any $\alpha\in \Delta^0_+$,
\begin{eqnarray}
&&S(E^\dagger_{\alpha+n\delta})=-q^{(\alpha,\alpha-2\rho)/2-n(\delta,\rho)}
F_{\alpha+n\delta}\nonumber\\
&&S(F^\dagger_{\alpha+n\delta})=-q^{-(\alpha,\alpha-2\rho)/2+n(\delta,\rho)}
E_{\alpha+n\delta}\nonumber\\
&&S(E^\dagger_{\delta-\alpha+n\delta})=-q^{(\delta-\alpha,\delta-\alpha-2\rho)
/2-n(\delta,\rho)}
F_{\delta-\alpha+n\delta}\nonumber\\
&&S(F^\dagger_{\delta-\alpha+n\delta})=-q^{-(\delta-\alpha,\delta-\alpha-2\rho)
/2+n(\delta,\rho)}
E_{\delta-\alpha+n\delta}\nonumber\\
&&S(\tilde{E}^{(i)\dagger}_{n\delta})=-q^{-n(\delta,\rho)}
\tilde{F}^{(i)}_{n\delta}\,.~~~~~
S(\tilde{F}^{(i)\dagger}_{n\delta})=-q^{n(\delta,\rho)}
\tilde{E}^{(i)}_{n\delta}\nonumber\\
&&S(E^{(i)\dagger}_{n\delta})=-q^{-n(\delta,\rho)}F^{(i)}_{n\delta}\,,~~~~~
S(F^{(i)\dagger}_{n\delta})=-q^{n(\delta,\rho)}E^{(i)}_{n\delta}
\end{eqnarray}
\vskip.1in
\noindent {\bf Proof:} The proof follows, from (\ref{ad1}),
(\ref{cartan-weyl3})
and lemma 4.3 and corollary 4.2, by induction in $n$.~~~$\Box$

Our main result is:
\vskip.1in
\noindent {\bf Theorem 4.4:} Every integrable highest weight module
$L(\Lambda)$
over $U_q({\cal G}^{(1)})$ corresponding to $q>0$ is equivalent to a unitary
module.
\vskip.1in
\noindent {\bf Proof:} The proof is similar to the one of theorem 4.2
for $U_q({sl(2)}^{(1)})$
case thanks to lemma 4.3, corollary 4.2, lemma 4.4 and
remark (\ref{remark}).~~~$\Box$
\vskip.1in
\noindent {\bf Corollary 4.3:} Every integrable highest weight module
$L(\Lambda
)$ over $U_q({\cal G})$ corresponding to $q>0$ is equivalent to a
unitary module.

\section{Concluding Remarks}
To summarize, in this paper we have investigated the complete reducibility and
unitarity of certain modules over the quantized affine Lie algebras. In our
proofs the Casimir operator (thus the universal $R$-matrix)
plays a key role and our approach is actually a
modest imitation of Kac's one\cite{Kac} and the one in \cite{Gould}
used in proving the similar results for the classical affine Lie algebras
and finite-dimensional simple Lie superalgebras, respectively.

The complete reducibility theorem 3.1 implies that the tensor product
$L(\Lambda)\otimes L(\Lambda')$ of the integrable irreducible highest
weight modules $L(\Lambda)$ and $L(\Lambda')$ is completely reducible and
the irreducible components are integrable highest weight representations.
This makes possible the computation of link polynomials
\cite{Reshetikhin}\cite{Zhang et al} associated with the quantized affine
Lie algebras. The point is that the universal $R$-matrix for the
quantized affine Lie algebras can be shown to satisfy the conjugation rule,
$R^\dagger=R^T$; therefore, braid generators are diagonalizable\cite{Gould}
on $L(\Lambda)\otimes L(\Lambda')$, regardless of multiplicity.
Our results also allow us to construct generalized Gelfand invariants
\cite{Gould et al} of the quantized affine Lie algebras. All details will
be reported in a separate publication.

\vskip.3in
\begin{center}
{\bf Acknowledgements:}
\end{center}
Y.Z.Z. would like to thank Anthony John Bracken for contineous encouragement
and suggestions, to thank Loriano Bonora for communication of preprint
\cite{KT} and to thank M.Scheunert for many patient explanations on quantum
groups during July and August of last year.
The financial support from  Australian Research Council
is gratefully acknowledged.

\end{document}